\documentclass[aps,pra,reprint,superscriptaddress,floatfix]{revtex4-2}
\usepackage{graphicx,tabularx,epstopdf,amsmath,bm}
\usepackage[utf8]{inputenc}
\usepackage[colorlinks=true,urlcolor=blue]{hyperref}
\usepackage{tcolorbox}

\begin{document}

\title{Phase-induced vortex pinning in rotating supersolid dipolar systems}

\author{Aitor Ala\~na}
\email{aitor.alana@ehu.eus}
\affiliation{Department of Physics, University of the Basque Country UPV/EHU, 48080 Bilbao, Spain}
\affiliation{EHU Quantum Center, University of the Basque Country UPV/EHU, 48940 Leioa, Biscay, Spain}
\author{Michele Modugno}
\email{michele.modugno@ehu.eus}
\affiliation{Department of Physics, University of the Basque Country UPV/EHU, 48080 Bilbao, Spain}
\affiliation{IKERBASQUE, Basque Foundation for Science, 48009 Bilbao, Spain}
\affiliation{EHU Quantum Center, University of the Basque Country UPV/EHU, 48940 Leioa, Biscay, Spain}

\author{Pablo Capuzzi}
\email{capuzzi@df.uba.ar}
\affiliation{Universidad de Buenos Aires, Facultad de Ciencias Exactas y Naturales,Departamento de Física. Buenos Aires, Argentina.}
\affiliation{CONICET - Universidad de Buenos Aires, Instituto de Física de Buenos Baires (IFIBA), Buenos Aires, Argentina.}
\author{D. M. Jezek}
\email{djezek@df.uba.ar}
\affiliation{CONICET - Universidad de Buenos Aires, Instituto de Física de Buenos Baires (IFIBA), Buenos Aires, Argentina.}
\date{\today}
\begin{abstract}
  We analyze the pinning of vortices for a stationary rotating dipolar
  supersolid along the low-density paths between droplets as a
  function of the rotation frequency. We restrict ourselves to the
  stationary configurations of vortices with the same symmetry as that
  of the array of droplets. In particular, such an analysis clearly
  reveals that vortices are not only pinned at local density minima,
  but instead their coordinates are smooth functions of the rotation
  frequency.  Our approach to explaining such a behavior exploits the
  fact that the wave function of each rotating droplet acquires a
  linear phase on the coordinates. Hence, the relative phases between
  the nearest neighboring droplets allow us to predict the position of
  the vortices in the intermediate low-density region. Here, we show
  that for a droplet distribution forming a triangular lattice, the
  phases of three neighboring droplets are needed for the correct
  description of the vortex location. In particular, for our confined
  system, we demonstrate that the estimate accurately reproduces the
  extended Gross-Pitaevskii results in the spatial regions where the
  neighboring droplets are well-defined.
\end{abstract}
\maketitle

\section{\label{sec:intro} Introduction}

Supersolids were experimentally created for the first time in 2017 in
spin-orbit coupled Bose-Einstein condensates (BECs)
\cite{li2017stripe}, BECs with cavity mediated interactions
\cite{Le17,Le17b}, and in 2019 in dipolar BECs \cite{Bo19,Ch19,Ta19},
with many other experiments featuring them afterwards
\cite{pollet2019,donner2019,Ta19b,Gu19,Na19,Ta21,He21,Pe21,No21,bottcher2021,sohmen2021,bia22,Bl22,sanchez2023heating}.
This state of matter combines the frictionless flow of the superfluids
with a translational symmetry breaking typical of crystals
\cite{Gross62,pitaevskii2016,Gross57,cristSov70,boni12,yukalov2020}. In
the case of dipolar supersolids, one can obtain them either by
generating a roton instability into an already condensed gas
\cite{Bo19,Ch19,Ta19,alana22,alana23,bia22,alana24,giovanazzi2002,Ro19,Ch18,poli23},
or by directly condensing the gas from a thermal cloud into a
supersolid \cite{sanchez2023heating}. Dipolar supersolids break the
translational symmetry by spontaneously forming a position-dependent
density distribution, which includes droplets of high density
separated by lower-density areas. In such a supersolid phase of
dipolar BECs \cite{roccu20,gal20,an20}, given that droplets are
separated by low density valleys, the barrier required for the
nucleation of vortices is reduced with respect to the superfluid case
(see e.g. \cite{casotti2024}).  In particular, for stationary rotating
systems, it was shown that low-density regions reduce the energetic
barrier for a vortex to enter the system, which lowers the nucleation
frequency and help in pinning the vortices in the interstitial zones
between droplets \cite{gal20}. For other long-range interacting
systems where droplets configurations are formed, the competition of
vortices to locate inside the superfluid droplets or at low-density
regions has also been theoretically analyzed in \cite{Henkel2012}.
Such a difference in the nucleation process has been early observed
experimentally in systems with contact interactions, as evinced by
comparing the results of, e.g, Abo-Shaeer \textit{et al.}
\cite{Abo-Shaeer2001}, where vortices are nucleated at the bulk, to
those of the experiment by R. A. Williams \textit {et al.}
\cite{williams10}, where a square lattice has been used.

The aim of this work is to predict the positions of vortices in a
stationary arrays in supersolid dipolar BEC \cite{roccu20,gal20,bia22}
forming a triangular lattice of droplets when it is subject to
rotation. Our approach consists in approximating the system wave
function through a superposition of the localized wave functions of
individual droplets.  Such a hypothesis is based on the fact that the
density is concentrated on the droplets, which are surrounded by very
low relative density valleys. Then, any droplet exhibiting axial
symmetry around a line parallel to the rotation axis acquires a
homogeneous velocity field \cite{rot20}, which is determined by the
velocity of the center of mass of the rotating droplet. In
consequence, the phase of the droplet wave function turns out to
exhibit a linear expression in terms of the spatial coordinates
\cite{rot20,je23}. Such an expression can be conveniently employed for
estimating the vortex positions between two neighboring droplets
through a simple formula, as it has been already shown for a BEC in
rotating square lattices \cite{je23}. In the present work, which
involves a triangular lattice, we show that the use of three
neighboring droplets in the model leads to very accurate values for
the vortex positions along the low-density region surrounded by such
droplets.

The paper is organized as follows.  In Sec. \ref{sec:distri} we
introduce the basic characteristics and parameters of a rotating
triangular lattice of droplets, which will be considered in our
analysis, and in Sec. \ref{sec:estima} we outline the method for
determining the vortex positions. In Sec. \ref{sec:stat} we describe
the confined system of dipolar atoms and show a typical stationary
configuration, whereas Sec. \ref{sec:nuclea} is devoted to the
determination of the coordinates of vortices of different
configurations. Finally, a summary of the results is given in
Sec. \ref{sec:summ}.

\section{\label{sec:distri} Triangular droplet lattice }

We start by considering the stationary configuration of a rotating
supersolid dipolar BEC, which forms an extended triangular lattice of
droplets. The key properties of this system are outlined below, and
will be then used to predict the characteristics of the vortex array
that emerges within the low-density regions between these droplets.
We assume the density is modulated as \cite{Pomeau1994,Zhang2019,Blakie2020,bia22}
\begin{equation}
\rho({\bf r} )= \rho_0 \, \left[ 1 + C \, \sum_{i=1}^3 \cos(\bm{q}_i\cdot\bm{r}) \right],
\label{eq:modelden}
\end{equation}
where the parameter $ C > 0$ represents the contrast. The vectors
  $ \bm{q}_i$, which lie in the $(x,y)$ plane, are defined by
\begin{equation}
\bm{q}_1 = q \hat{y} \, , \, \, 
\bm{q}_2 = - \frac{1}{2} q \hat{y} + \frac{\sqrt{3}}{2} q \hat{x} \, , \, \,
\bm{q}_3 = - \frac{1}{2} q \hat{y} - \frac{\sqrt{3}}{2} q \hat{x}\, ,
\end{equation}
with $q= 2 \pi /\lambda$. It is worth noting that in a realistic
setup, the overall density factor may exhibit a dependence on the
coordinate $z$, $\rho_0=\rho_0(z)$. This dependence can be modeled by
a Gaussian or Thomas-Fermi distribution. However, for the purposes of
the subsequent discussion, this dependency can be safely disregarded
without loss of generality.

%%Figure 1
\begin{figure}[t]
\includegraphics[width=1\columnwidth]{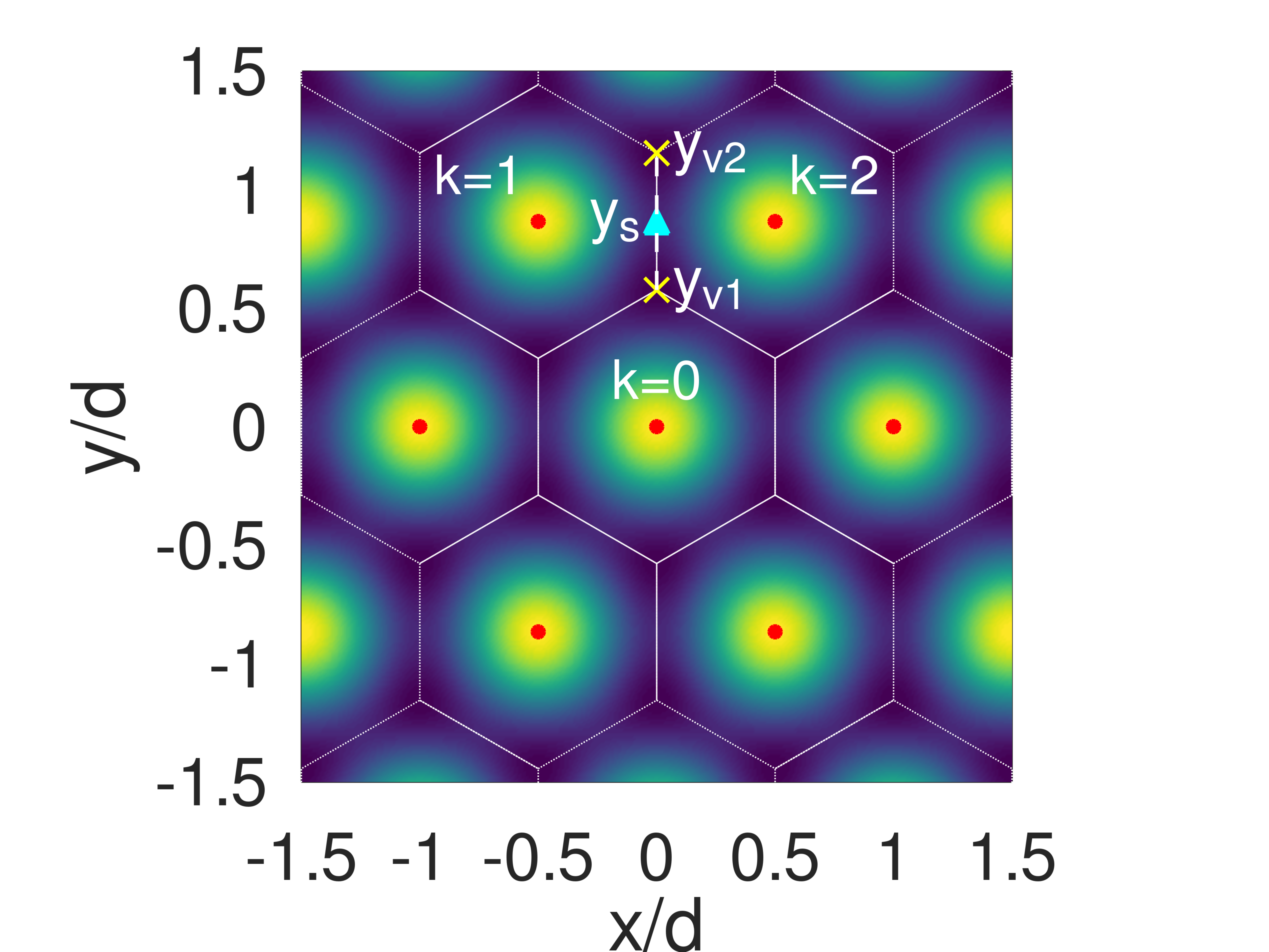}%
\caption{\label{fig:den} Density plot of the density distribution
  given by Eq.~(\ref{eq:modelden}), where the density maxima are
  marked with red dots. The thin white lines indicate the low-density
  paths around the droplets, while the dashed line marks the specific
  path described in the text. In such a path, the saddle point at
  $y_s=\lambda$ is marked with a triangle, and the minima at
  $y_{v1}= d / \sqrt{3}$ and $y_{v2}= 2d / \sqrt{3}$ (vertices) with
  crosses.  }
\end{figure}
In Fig. \ref{fig:den} we show a plot of the density distribution in
Eq. \eqref{eq:modelden}. By analyzing its maxima, minima, and saddle
points, we can characterize the density pattern as follows. The
distance between neighboring droplet maxima is
$d = 2 \lambda/ \sqrt{3}$. The minima are located at equidistant
positions from three neighboring droplets, namely at the
\textit{vertices} of the hexagonal structure depicted in the
figure. In the following, we will use the term \textit{path} to denote
the line segments connecting them (namely, each side of the
hexagon). When the droplets have all the same size and shape, as in
the ideal case depicted in Fig. \ref{fig:den}, all paths in
between the first ring of droplets are equivalent. These are the
paths that will be specifically relevant for the discussion in
Sec. \ref{sec:stat}. 
Therefore, here we will focus ourselves on the vertical
path marked by the dashed line in Fig. \ref{fig:den}, without loss of
generality. For this specific case, the two vertices are located along
the $y$-axis at $y_{v1}= d / \sqrt{3}$ and $y_{v2}= 2 d /
\sqrt{3}$. At the center of the path, $y_s=\sqrt{3}/2 d$, which
corresponds to the middle point between the two neighboring droplets,
the density displays a saddle point.

When the supersolid lattice is put under rotation, stationary vortices
will appear along the low-density paths between the droplets
\cite{je23,casotti2024,poli23}. The position of the vortex along those
paths, denoted as $Y_v$ for the specific path considered above, can be
easily estimated using the ansatz discussed below, in
Sec. \ref{sec:estima}.

In Sec. \ref{sec:stat} we will consider a finite realization of this
system, which can be achieved through numerical calculations. In order
to do so, we will introduce a harmonic trap to confine the
system. Given that we also subject the system to rotation at constant
frequency $\Omega$ along the $z$-axis, the effective confinement
varies with $\Omega$. Then, the distribution of droplets and their
densities vary as well. We will verify that a number of droplets
arrange in a triangular lattice, and hence compute $d$ and the
remaining geometrical quantities from the obtained densities for each
frequency.

\subsection{\label{sec:estima} Estimate of the vortex positions }

In this section, we outline the way to estimate the position of the
vortices, following Ref. \cite{je23}.  We assign to each droplet $k$ a
localized wave function $ w_{k} ({ \bm{r}, \Omega })$ normalized
to unity, where $\bm{r}=(x, y, z)$. Hence, the wave function of the system of droplets
can be approximated by
\begin{equation}
\psi_D( \bm{r},t )= \sum_{k} w_{k} ({ \bm{r}, \Omega })
 \, e^{i \phi_{k} (t)} \sqrt{N_{k}(t)} \,,
\label{orderp}
\end{equation}
where $ {N_{k}(t)}$ is the number of particles of the droplet, $ \phi_{k} (t)$ its global phase, 
and the indices $k$ runs upon all the droplets. 
Given the axial symmetry of each droplet, we may further 
 approximate \cite{rot20}
\begin{equation}
 w_{k}(\bm{r},\Omega) = |w_{k}(\bm{r},\Omega)| e^{i \frac{m}{\hbar}
(\bm{r}-\bm{r}_{\text{cm}}^{k})\cdot(\bm{\Omega}\times\bm{r}_{\text{cm}}^{k})},
 \label{eq:localg}
\end{equation}
where we have fixed to zero the phase of $w_{k}$ at the center of mass
of the droplet.

\textit{Two-droplet case.}  Let us first consider the case in which
the vortex sits between two neighboring droplets, labeled as $k'$ and
$k$. Specifically, we examine the two droplets indicated in the upper
section of Fig. \ref{fig:den}, which are symmetric with respect to the
vertical $y$-axis. We denote the generic coordinates of a vortex core
in the $z=0$ plane as $(X_v, Y_v, 0)$. Due to symmetry, a vortex lying
between these two droplets will have $X_v=0$, while the vertical
coordinate $Y_v$ can be obtained by requiring the vanishing of the
wave function at the vortex core, $\psi_D(X_v, Y_v,0)=0$, namely
\begin{equation}
 w_{k'}( \bm{r})
 \, e^{i \phi_{k'}} \sqrt{N_{k'}} + w_k(\bm{r})
 \, e^{i \phi_k} \sqrt{N_k} = 0 \, ,
\label{zeron}
\end{equation}
where we have omitted the time dependence for ease of notation.
By writing,
\begin{multline}
(\bm{r}-\bm{r}_{\text{cm}}^k)\cdot(\bm{\Omega}\times\bm{r}_{\text{cm}}^k) =
(\bm{r}-\bm{r}_{\text{cm}}^{k'})\cdot(\bm{\Omega}
\times\bm{r}_{\text{cm}}^{k'}) \\ + \bm{r} \cdot (\bm{\Omega} \times (\bm{r}_{\text{cm}}^k - \bm{r}_{\text{cm}}^{k'})),
 \label{phaseg}
\end{multline}
 Eq. (\ref{zeron}) can be rewritten as,
\begin{equation}
 \sqrt{N_k}|w_k| \, e^{i (\frac{m }{\hbar} \bm{r} \cdot (\bm{\Omega} \times (\bm{r}_{\text{cm}}^k - \bm{r}_{\text{cm}}^{k'}) ) - \varphi_k ) } + 
 \sqrt{N_{k'}} |w_{k'}| \, = 0 \,,
\label{exc3}
\end{equation}
where $\varphi_k(t)= \phi_{k'}(t) -\phi_k(t) $ is the phase
difference between the centers of such neighboring droplets.

In terms of the center-of-mass coordinates one has,
\begin{multline}
  \bm{r} \cdot (\bm{\Omega} \times (\bm{r}_{\text{cm}}^k -
  \bm{r}_{\text{cm}}^{k'}) ) = -x (
  y^{k}_{\text{cm}}-y^{k'}_{\text{cm}} ) \Omega \\+ y (
  x^{k}_{\text{cm}}-x^{k'}_{\text{cm}} ) \Omega \,.
\label{corden}
\end{multline}
As for the droplet label $k$, here we set $k=0$ for the central
droplet, and let run $k$ clockwise for the outer droplets, as
indicated in Fig. \ref{fig:den}. Then, considering the case of the two
droplets with $k=2$ and $k'=1$ in Eq. (\ref{corden}), for which
$ y^{1}_{\text{cm}} = y^{2}_{\text{cm}} $, we may obtain, from the
condition that the imaginary and real parts of Eq. (\ref{exc3}) should
vanish, the vortex coordinate $Y_v(t)$,
\begin{equation}
Y_{v}(t)= \left( \frac{ \varphi(t)}{\pi} + 
2l + 1 \right) \, \frac{ \pi \hbar}{ m d \Omega} \,,
\label{eq:vortimel}
\end{equation}
where $d=x^{2}_{\text{cm}}-x^{1}_{\text{cm}} $ is the distance between
the center of mass of the droplets, and $\varphi=\varphi_2= \phi_1-\phi_2$. Here, $l$
is an integer number labeling different possible solutions, with $l=0$
corresponding to the first vortex that enters through that path
\cite{je23}.  It is also important to remark that the coordinates of
the center of mass increase as functions of the rotation frequency due
to the centrifugal force, and we will estimate its position by
searching the density maxima of the droplets.

\textit{Three-droplet case.}  In principle, in a triangular lattice,
when the location of a vortex is near a vertex of the droplet lattice,
the presence of a third neighboring droplet should affect the vortex
position, and hence it becomes important to take such an effect into
account. Then, one can approximate the wave function in such a region
as
\begin{equation}
\psi_D( \bm{r},t ) \simeq \sum_{k=0}^2 |w_{k}(\bm{r},\Omega)| e^{i \frac{m}{\hbar}
(\bm{r}-\bm{r}_{\text{cm}}^{k})\cdot(\bm{\Omega}\times\bm{r}_{\text{cm}}^{k})+i \phi_{k} }
 \sqrt{N_{k}} \,.
\label{orderpv}
\end{equation}
In this case we cannot extract an analytical expression for the vortex
coordinates. However, by adequately modelling the individual wave
functions of the droplets, an approximate solution can be obtained.
Here we shall consider the two droplets at the first ring ($k=1,2$)
together with the central one ($k=0$), a case that will be relevant
for the finite realization presented in the following section.  For
such a purpose we will approximate $ |w_{k}| $ by Gaussian functions
with widths $a$ and heights which almost reproduce the characteristics
of our droplets. We further assume that $ \phi_{k} = \phi_0$ for all
sites.  With these approximations, we can again obtain an expression
for $Y_v$ by imposing that the wave function of Eq. (\ref{orderpv})
vanishes at the position of the vortex core.  In particular, the value
of $Y_v$ is given by the solution of
\begin{equation}
\sqrt{\frac{N_0}{N_1}}\, e^{\dfrac{d (d - \sqrt{3} Y_v)}{2 a^2}} + 2 \cos\left(\frac{m d}{2\hbar} \Omega Y_v\right)=0,
\label{eq:threedroplets}
\end{equation}
where we have accounted for the fact that in a finite realization the
central droplet population $N_0$ may be different from the population
$N_1=N_2$ of the other two droplets.  Notice that
Eq. (\ref{eq:threedroplets}) has multiple solutions which are related
to those labeled by $l$ in Eq. (\ref{eq:vortimel}).

The above Eq. \eqref{eq:threedroplets}, along with
Eq. \eqref{eq:vortimel}, constitutes one of the central results of the
present work. In Sec. \ref{sec:nuclea}, we will compare it with exact
results from numerical simulations, demonstrating its accuracy in
predicting vortex positions.

\section{\label{sec:stat} Rotating stationary supersolid}

In order to present a practical case study, we focus on investigating
a rotating stationary supersolid configuration within a dipolar system
akin to the one studied in Ref. \cite{gal20}.  Specifically, we
consider a Bose gas composed by $N=1.1\times10^{5}$ dipolar $^{162}$Dy
atoms trapped by an axially symmetric harmonic trap of frequencies
$\{\omega_r,\omega_z\}= 2 \pi \times \{60,120\}$ Hz.  For this atomic
species, the dipolar scattering length is $a_{dd}=130a_0$, where $a_0$
stands for Bohr radius.  The magnetic dipoles are considered to be
aligned along the $z$ direction by a magnetic field $\bm{B}$. The
$s$-wave scattering length of the contact interaction is fixed to
$a_{s}=92a_0$ throughout the whole paper. The system is set to rotate
at an angular velocity $\Omega$ around the polarization axis.

The advantage of this specific configuration is that it features a
triangular supersolid lattice as the ground state, which is the
closest packing configuration and thus is of special
interest. However, the model developed in this paper does not require
any specific geometry and could be applied to other supersolid
configurations as long as the positions of the droplets are correctly
taken into account\footnote{We perform an extension from the 2 droplet
  model to a one featuring 3 droplets, which is convenient for the
  specific case. One should be able to use the same model for other
  configurations since it is easily adaptable.}.

 We consider the gas to be at $T=0$, thus no thermal fluctuations are
 taken into account. We describe the system using the usual extended
 Gross Pitaevskii (eGP) theory, which includes both the quantum
 fluctuations in the form of the Lee-Huang-Yang (LHY) correction
 \cite{schutzhold2006,Li12,wachtler2016,schmitt2016} and the dipole-dipole
 interaction \cite{ronen2006}. To account for the rotation of the
 condensate we will work in the rotating frame, for which an
 additional term is introduced into the energy functional
 \cite{castin1999,modugno2003}. The energy functional of such a system
 can be written as
 $E_{\text{GP}}+E_{\text{dd}}+E_{\text{LHY}}+E_{\Omega}$, with
\begin{align}
E_{\text{GP}} &= 
\int \left[\frac{\hbar^2}{2m}|\nabla \psi(\bm{r})|^2 + V(\bm{r})n(\bm{r})+\frac{g}{2} n^2(\bm{r})
\right]d\bm{r}\,,
\nonumber\\
E_{\text{dd}} &=\frac{C_{\text{dd}}}{2}\iint n(\bm{r})V_{\text{dd}}(\bm{r}-\bm{r}')n(\bm{r}') d\bm{r}d\bm{r}'\,,
\nonumber\\
E_{\text{LHY}} &=\frac{2}{5}\gamma_{\text{LHY}}\int n^{5/2}(\bm{r})d\bm{r}\,,
\label{eq:GPenergy}\\
E_{\Omega} &= -\Omega\int \psi^{*}(\bm{r})\hat{L}_{z}\psi(\bm{r})d\bm{r}\,,
\nonumber
\end{align}
where $E_{\text{GP}}=E_{\text{k}}+E_{\text{ho}}+E_{\text{\text{int}}}$
is the standard GP energy functional including the kinetic, potential,
and contact interaction terms,
$V(\bm{r})=(m/2)\sum_{\alpha=x,y,z}\omega_{\alpha}^{2}r_{\alpha}^{2}$
is the harmonic trapping potential, and $g=4\pi\hbar^2 a_{s}/m$ is the
contact interaction strength. The system wave function $\psi(\bm{r})$
is normalized to the total number of particles $N$ and the condensate
density is given by $n(\bm{r})=|\psi(\bm{r})|^2$.  The inter-particle
dipole-dipole potential is
$V_{\text{dd}}(\bm{r})= (1-3\cos^{2}\theta)/(4\pi r^{3})$ with
$C_{\text{dd}}\equiv\mu_{0}\mu^2$ its strength, $\mu$ the modulus of
the dipole moment $\bm{\mu}$, $\bm{r}$ the distance between the
dipoles, and $\theta$ the angle between the vector $\bm{r}$ and the
dipole axis, $\cos\theta=\bm{\mu}\cdot\bm{r}/(\mu r)$. The LHY
coefficient is
$\gamma_{\text{LHY}}={128\sqrt{\pi}}{\hbar^{2}a_s^{5/2}}/(3m)\left(1 +
  3\epsilon_{\text{dd}}^{2}/2\right)$, with
$\epsilon_{\text{dd}}=\mu_0 \mu^2 N/(3g)$.  The last term $E_{\Omega}$
accounts for the rotating frame, with
$\hat{L}_{z}=-i\hbar(x\partial_y -y\partial_x)$ representing the
angular momentum operator along $z$.
To obtain the supersolid stationary states in the rotating frame, we
perform numerical simulations \footnote{For the numerical simulations
  we use a computation box of
  $10\mu$m$\times$$10\mu$m$\times$$12\mu$m with a grid
  $\{ 256, 256, 64\}$ } in which we minimize the above energy
functional employing a conjugate gradient method (see
e.g. \cite{press07}).  Among the several possible stationary
  configurations, we select those corresponding to a triangular
  supersolid lattice by means of a suitable choice of the trial wave
  function \footnote{It should be noted that the conjugate gradient
    approach employed to minimize the eGP energy functional inherently
    yields local minima. In this context, employing different trial
    wave functions can generate alternative lattice geometry
    configurations that are nearly degenerate in energy. In the
    present case, we have numerically verified that the triangular lattice
    configuration indeed corresponds to the minimal energy solution,
    i.e., the ground state, within the range of rotation frequencies
    considered here.}.

\begin{figure}[t]
 %\flushleft
 \centerline{\includegraphics[width=0.74\columnwidth]{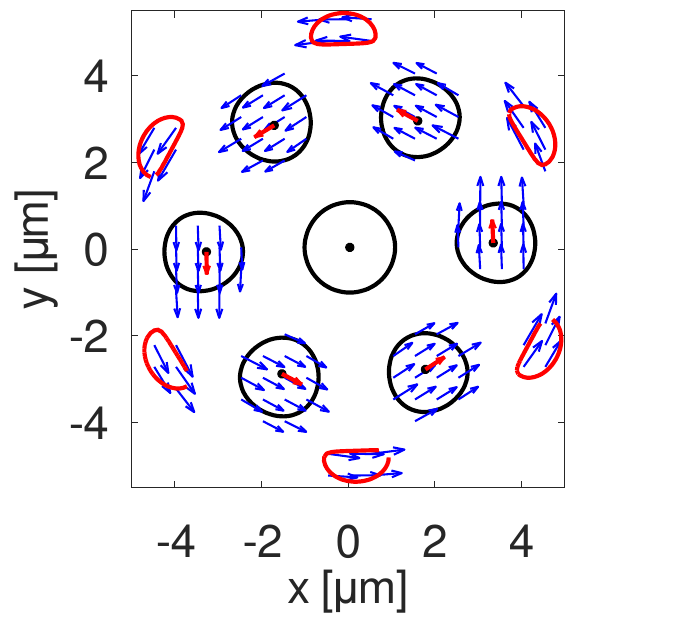}}
 \centerline{\hspace{0.9cm}~\includegraphics[width=0.95\columnwidth]{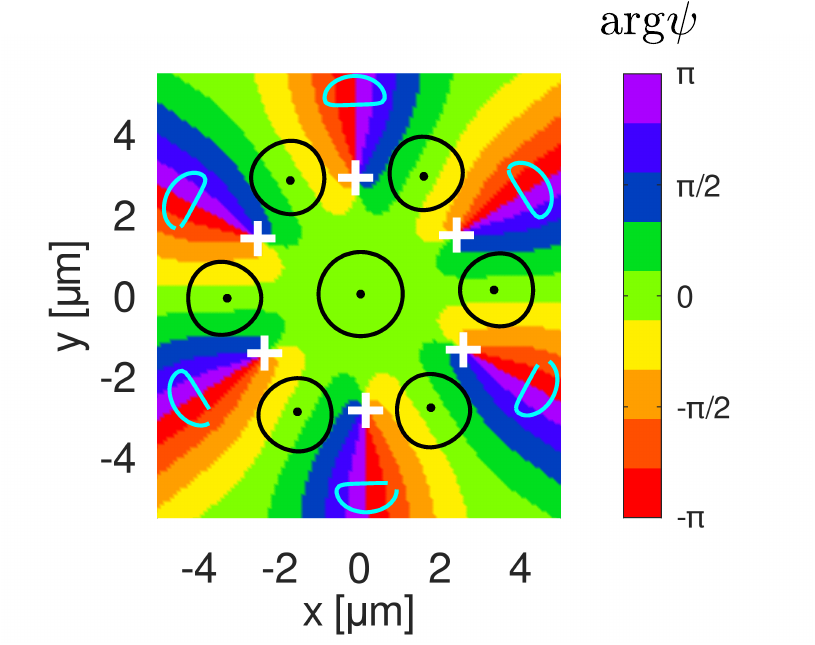}}
 \caption{Typical density and phase configuration of the stationary
   supersolid triangular lattice obtained from the eGP simulations, in
   the rotating frame. Here $\Omega=2\pi\times20$ Hz.  (top)
   Isodensity contours and velocity field around the stationary
   droplets.  The black curves correspond to the contours around the
   six first-ring droplets at a density value of
   $0.2\rho_{\text{max}}$, and the red curves around the low-density
   clouds at a density value of $0.04\rho_{\text{max}}$, where
   $\rho_\text{max}$ is the maximum density value.  The red arrows at
   the droplet maxima correspond to the $\bm{v}_k$.  (bottom) Color
   map of the phase of the supersolid wave function at the plane
   $z=0$.  The plus signs mark the location of the vortices, and the
   curves are isodensity contours of the droplets for the same values
   as in the top panel. }
 \label{fig:rhovx}
\end{figure}

A typical configuration is displayed in Fig. \ref{fig:rhovx},
featuring the density contours and velocity field in the upper panel,
and the phase distribution along with the position of the vortex cores
in the lower panel. This figure corresponds to the case with
$\Omega=2\pi\times20$ Hz, and serves as a representative illustration
for all cases within the range of rotation frequencies considered in
this work.  Figure \ref{fig:rhovx} reveals well-localized, circularly
symmetric densities of the gas droplets (depicted in black) arranged
in a triangular structure formed by a central droplet at
$(x, y) = (0, 0)$, and six droplets located along a
ring around it. It may be seen that each of these droplets exhibits a uniform
velocity field $\bm{v}_k=\bm{\Omega}\times
\bm{r}^k_{\text{cm}}$. Additionally, at the border, very low-density
clouds with non-circular shapes (in red) are present and display a
diffuse distribution with an extended velocity field.  Then, such a
cloud is not included in the region where $\psi_D$ is defined. Between
the droplets, we observe the presence of vortices whose positions form
a lattice structure determined by the periodic arrangement of the
supersolid droplets, see bottom panel in Fig. \ref{fig:rhovx}.

\subsection{\label{sec:nuclea}Vortex pinning}

\begin{figure}[t]
 \includegraphics[width=0.9\columnwidth]{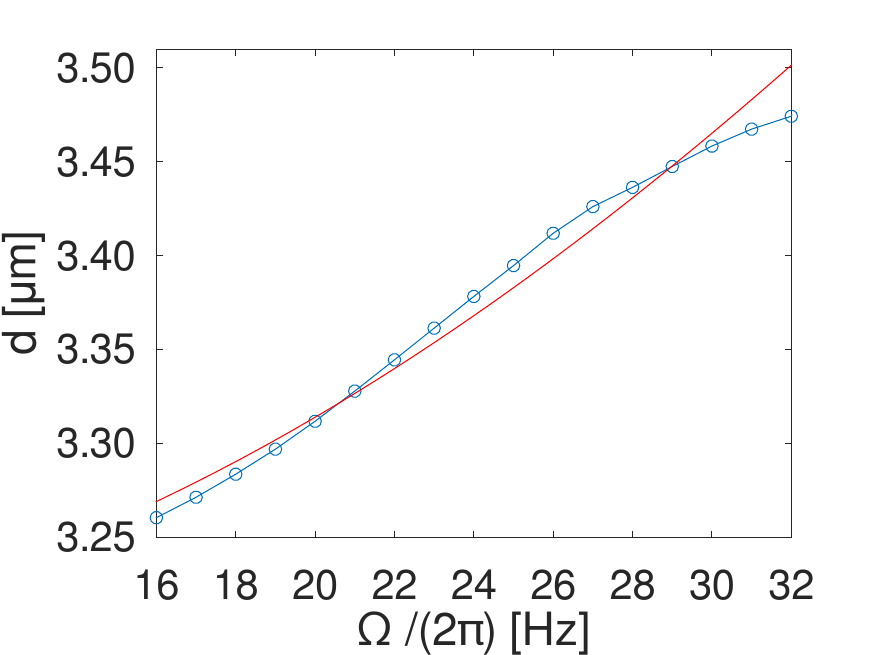}%
 \caption{\label{fig:dmedia} Mean value of the inter-droplet distance
   $d$ (blue circles) as a function of the rotation frequency
   $\Omega$.  The data have been obtained by identifying the density
   maxima from the calculated supersolid configurations. The solid red
   line correspond to the distance for a non-rotating configuration
   with an effective trap frequency
   $\tilde{\omega}_r= \sqrt{\omega_r^2-\Omega^2}$.}
\end{figure}

In order to investigate the position of the vortices as a function of
the rotation, we employ the method outlined in
Sec. \ref{sec:estima}. Here we focus on the location of the vortices
along the low-density paths bounded by two vertices, such as the line
joining $y_{v1}$ and $y_{v2}$ in Fig. \ref{fig:den}. We begin by
considering the effect of pairs of neighboring droplets. The mean
relative distance $d$ between droplet pairs is shown in
Fig. \ref{fig:dmedia} as a function of the rotation frequency $\Omega$
(blue circles).  We observe an increase of such a distance with the
frequency, which can be mainly attributed to the effect of the
centrifugal force acting on the particles. This can be proved by
comparison with the inter-droplet distance of a non-rotating gas
trapped at the effective frequency
$\tilde{\omega}_r=\sqrt{\omega_r^2-\Omega^2}$ so as to mimic the
centrifugal force effect, shown in the same Fig. \ref{fig:dmedia}
(solid red line).

\begin{figure}[!ht]
\centering
\includegraphics[width=0.9\columnwidth]{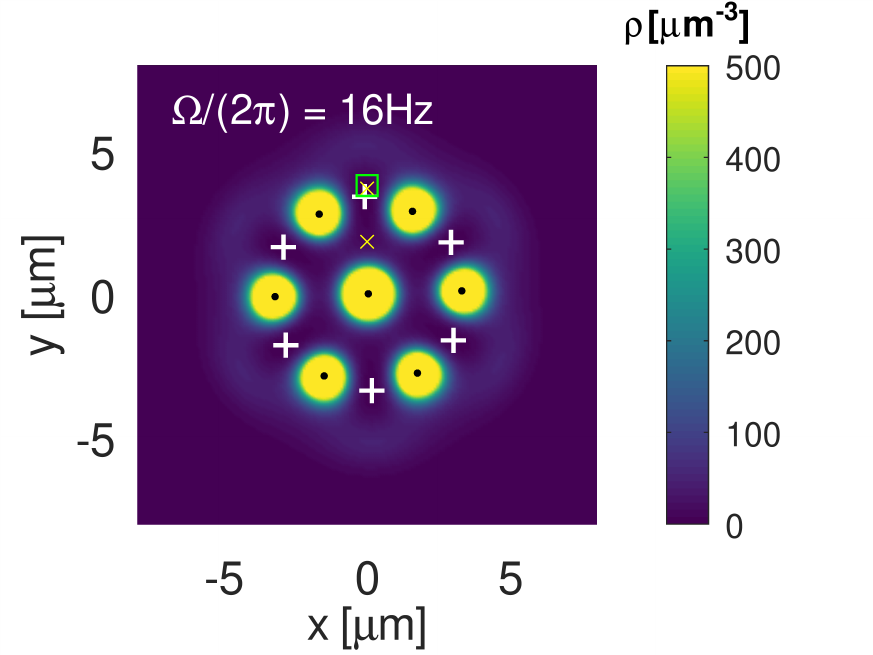}\\
\includegraphics[width=0.9\columnwidth]{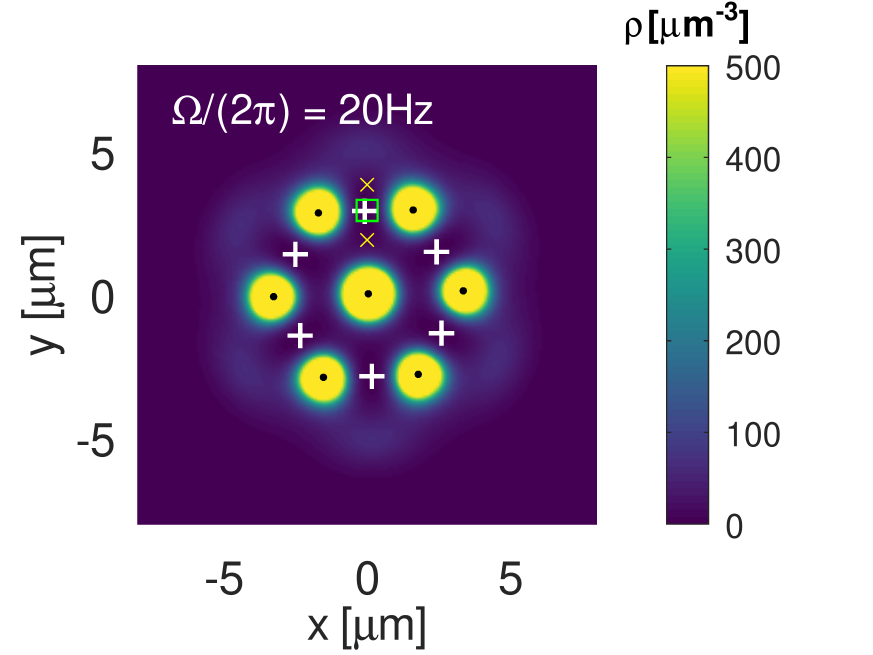}\\
\includegraphics[width=0.9\columnwidth]{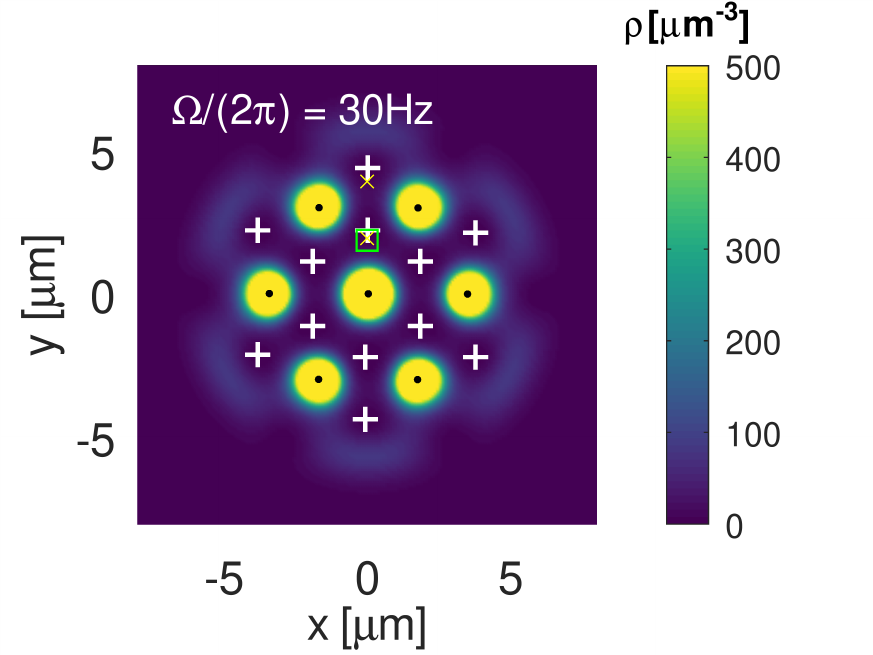}
\caption{\label{fig:rhos} Density plots of stationary supersolid
  configurations obtained from the eGP simulations at the $z=0$ plane
  for some representative rotation frequencies:
  $\Omega/(2\pi)=16,\,20,$ and $30$ Hz (from top to bottom,
  respectively).  The black dots mark the maxima of the density.  The
  vortex positions determined with the plaquette method are shown as
  white plus symbols ($+$).  The position of the vortex core obtained
  from the analytical ansatz, along the straight path between the
  vertices $y_{v1}$ and $y_{v2}$ [yellow crosses ($\times$)], is
  indicated by a green square.  }
\end{figure}

Then, we extract the positions of any vortices present in the system
using a plaquette method \cite{Foster2010} and compare them to our
estimate in Eq. \eqref{eq:vortimel}.  In Fig. \ref{fig:rhos} we show
examples of the stationary density distribution at different rotation
frequencies, along with the vortex locations.  This figure shows that
the vortices are not necessarily pinned to the vertices, but rather
along the low-density paths that connect them.  The configurations
conserve the triangular symmetry both for the density and phase
profiles, and we observe that they display vanishing phase differences
among droplet centers, i.e., $ \varphi_k =0, \forall k$. This leads to
the prediction $Y_v= (2 l+1)\pi\hbar/(md\Omega)$.  It is worth
noting that the position of the vortex along the straight path between
the vertices $y_{v1}$ and $y_{v2}$ (see figure) corresponds to the
solution with $l=0$, whereas different values of $l$ identify
additional vortices that may enter the system from outside.
Nevertheless, we remark that those additional vortices that appear,
e.g., for $\Omega/(2\pi)=30$ Hz (bottom panel of Fig. \ref{fig:rhos})
cannot be properly described by the ansätze \eqref{eq:vortimel} and
\eqref{eq:threedroplets} because they are nucleated in the low-density
cloud, outside the region of validity of the analytical approach.

\begin{figure}
 \includegraphics[width=1 \columnwidth]{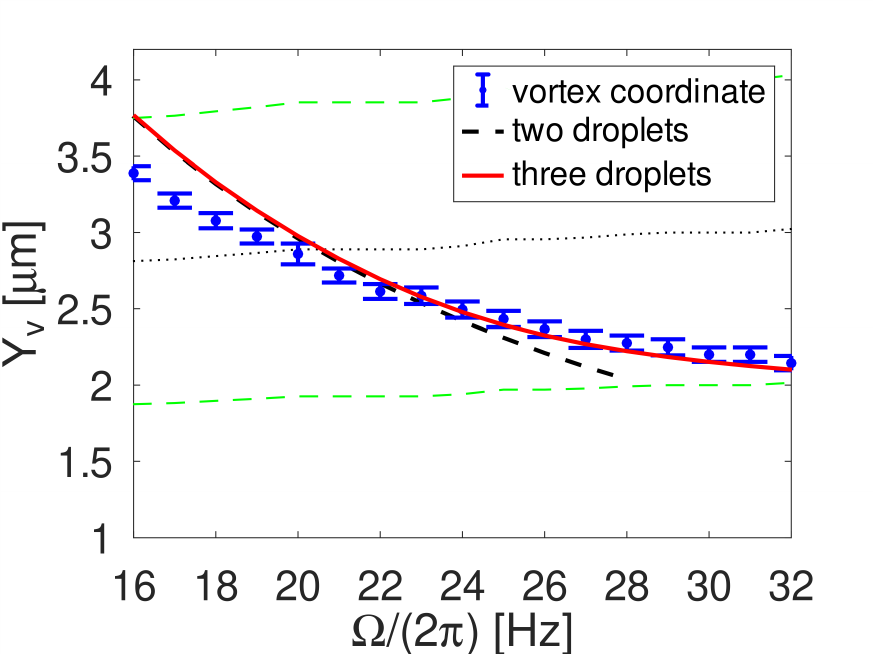}%
 \caption{\label{fig:vorttri} Vortex coordinate $Y_v$ as a function of
   the rotation frequency $\Omega$, for the stationary configurations
   and along the straight path with $X_v=0$ (see Fig. \ref{fig:den}).
   The coordinates are extracted from full eGP simulations using a
   plaquette method (see text) are marked as blue points with their
   corresponding error bar.  The prediction provided by
   Eq. (\ref{eq:vortimel}) with $l=0$ (black dashed line) and
   \eqref{eq:threedroplets} (solid red line) have been calculated
   using the values of $d$ obtained from the numerical simulations,
   which are shown in Fig. \ref{fig:dmedia}.  They correspond to the
   ansatz based on two and three droplets, respectively. The
   parameters of the three-droplet ansatz are $a=0.75\mu$m and
   $\sqrt{N_0/N_1}=1.06$. The (green) dashed lines represent positions
   of the vertices, whereas the (black) dotted line indicates the
   position of the saddle point. }
\end{figure}
 
At this point, we are now able to compare the analytical predictions
of Eqs. \eqref{eq:vortimel} and \eqref{eq:threedroplets} with the
extracted values of the vortex coordinates $Y_v$ along the path
indicated in Fig. \ref{fig:den} as a function of the
rotating frequency, as summarized in Fig. \ref{fig:vorttri}. Overall,
this figure demonstrates that the analytical ansatz discussed in
Sec. \ref{sec:estima} provides an accurate prediction for the
positions of the vortex cores between supersolid droplets in
stationary rotating configurations.  It is also worth noting that the
pinning at the saddle points, represented by a dotted line in the
graph, and density minima do not seem to be favored with respect to
other points along such paths, as stated previously (see
e.g. \cite{poli23}).  Instead, the vortex position smoothly changes as
a function of the rotation frequency.  As a matter of fact, in a
rotating supersolid, the slow variation of the vortex location arises
from the imprinted velocity field on the droplets, rather than from
density holes that typically pin vortices in non-rotating systems.

Let us now comment about the two- and three-neighboring-droplet
approximations.  Although the two-droplet ansatz is not expected to
hold far from the saddle point, where the third droplet effect should
be taken into account, the ansatz accurately predicts the rotation
frequency at which vortices locate near such a point, through an
analytical formula. Instead, the three-droplet model permits us to
numerically estimate with accuracy the position of vortices along the
line joining the saddle point and the vertex $y_{v1}$. We note that in
such a region the three neighboring droplets are well defined. This is
evident from Fig. \ref{fig:vorttri}, where the positions of the
vertices are indicated by (green) dashed lines. We may first mention
that the predicted rotation frequency for the presence of a vortex at
the saddle $y_s$, using the two- and three-droplet approximations
differ in less than 0.3\%, as the exponential term in Eq.\
(\ref{eq:threedroplets}) is smaller than $10^{-2}$. Notice that
neglecting such an exponential term altogether, both approaches
coincide.  In contrast, for the vertex $y_{v1}$, the two-droplet
rotation frequency estimate is given by
$\Omega/2 \pi = { \hbar \sqrt{3}}/{( 2 m d^2 )} \simeq 28$ Hz;
whereas, using the three-droplet approach of
Eq. (\ref{eq:threedroplets}), one obtains $\Omega/2\pi\simeq 37$ Hz.
The last result may be easily obtained by considering that the
argument of the exponential vanishes at the vertex $y_{v1}$, and hence
the frequency at which the vortex should reach such a vertex satisfies
\begin{equation}
\cos\left( \frac{m d^2 \Omega }{2\hbar\sqrt{3}}\right) = -\frac{1}{2}\sqrt{\frac{N_1}{N_0}}.
\end{equation}
Then, assuming equal populations $N_0=N_1$ the above equation leads to
$\Omega/2\pi = 2\hbar/(\sqrt{3}md^2)\simeq 37$ Hz for the lowest
$\Omega$ solution, which improves the value of the vortex coordinate
with $l=0$ of Eq. \eqref{eq:vortimel}, consistently with the numerical
findings of the eGP simulations. In summary, we have shown that the
three-droplet model better describes the vortex position as a function
of rotation frequency between the saddle $y_s$ and vertex $y_{v1}$.
Moreover, as seen from the previous analysis, the inclusion of the
third droplet explains the fact that for a given frequency it is more
likely to find the vortex near the vertex than in the proximity of the
saddle.

We finally note that we do not apply the three-droplet approach in the
line from the saddle point to the vertex $y_{v2}$ since near such a
vertex the formation of a third neighboring droplet is not
observed. The deviation from the estimate there observed, see
Fig. \ref{fig:vorttri}, can be caused by the presence of what we call the
cloud which, as shown in Fig. \ref{fig:rhovx}, has a very low density
and a different symmetry.

\section{\label{sec:summ}Summary and concluding remarks}

We have shown that when a dipolar supersolid is subjected to rotation,
the positions of vortices between two neighboring droplets can be
predicted in terms of the rotation frequency and the inter-droplet
distance. Such a distance can be roughly estimated using the
non-rotating system with a harmonic potential that mimics the net
confinement produced by the rotating trap. The vortex positions are a
smooth function of the rotation frequency and are distributed along
the low-density paths between the droplets, instead of being fixed at
a density minimum. Such a formulation applies in the regions where
robust droplets acquire on-site axially symmetric profiles. We have
further shown that a very accurate value of the vortex locations can
be numerically obtained by considering three neighboring droplets
within the triangular lattice.  In the present case, due to the
external confinement, three well-formed neighboring droplets could be
observed only around the first vertex, but given that our estimate
remains valid from the vertex up to the saddle, we may conclude that
the model should work well for less confined systems where more
droplets are formed around other vertices of the triangular lattice.

As a final remark, the approach can be generalized to more complex
droplets configurations as long as the droplets themselves are axially
symmetric and well defined. Vortices will likely be placed in areas
where two or three neighboring droplets are enough to precisely
predict their positions, regardless of the lattice structure of the
supersolid.

\begin{acknowledgments}
We acknowledge fruitful discussions with I. L. Egusquiza.
This work was supported by Grant PID2021-126273NB-I00 funded by MCIN/AEI/
10.13039/501100011033 and by ``ERDF A way of making Europe'', by the Basque
Government through Grant No. IT1470-22, and by the European Research
Council through the Advanced Grant ``Supersolids'' (No. 101055319). P.C.
acknowledges support from CONICET and Universidad de Buenos Aires, through grants 
PIP 11220210100821CO and UBACyT 20020220100069BA, respectively.
\end{acknowledgments}

\end{document}